\documentclass{ifacconf}

\usepackage{savesym}
\savesymbol{AND}

\usepackage{url}
\usepackage{amsmath}
\usepackage{amssymb}
\usepackage{graphicx}
\usepackage{algpseudocode}
\usepackage{algorithm}
\usepackage{amsmath}
\usepackage{tikz}
\usepackage{epstopdf}
\usepackage{mathtools}
\usepackage{natbib}

\newtheorem{proposition}{Proposition}
\newtheorem{definition}{Definition}
\newtheorem{lemma}{Lemma}
\newtheorem{remark}{Remark}

\newtheorem{assumption}{Assumption}

\renewcommand{\vec}{\mathrm{vec}}

\DeclareMathOperator{\lcm}{lcm}
\DeclareMathOperator{\modd}{mod}

\usepackage{bbm}

\definecolor{color1}{rgb}{0.7882,0.5804,0.7804}
\definecolor{color2}{rgb}{0.5686,0,0.2471}

\DeclareRobustCommand\comm{\raisebox{-1.4pt}{\tikz{\draw[-,color2,solid,line width = 1.2pt](0,0) -- (8mm,0);\node[color1,line width = 1.2pt,minimum size=1.9mm,inner sep=0pt] at (.4,0) {\color{color2}\large$\times$};}}}

\DeclareRobustCommand\comp{\raisebox{0pt}{\tikz{\draw[-,color1,solid,line width = 1.2pt](0,0) -- (8mm,0);\node[circle,color1,line width = 1.2pt,minimum size=1.9mm,draw,inner sep=0pt] at (.4,0) {};}}}

\renewcommand{\footnoterule}{
  \kern -2pt
  \hrule width 0.3\textwidth height .5pt
  \kern 2pt
}

\renewcommand{\cite}{\citep}

\begin{document}

\begin{frontmatter}
\title{Private and Secure Coordination of Match-Making for \\ Heavy-Duty Vehicle Platooning\thanksref{footnoteinfo}}

\thanks[footnoteinfo]{The work of F. Farokhi was supported by a McKenzie Fellowship and an Early Career Grant from Melbourne School of Engineering. The work of K. H. Johansson was supported by Knut och Alice Wallenbergs Foundation (KAW), Swedish Foundation for Strategic Research (SSF), and Swedish Research Council (VR).}

\author[First]{Farhad Farokhi}
\author[First]{Iman Shames}
\author[Second]{Karl H. Johansson}

\address[First]{Department of Electrical and Electronic Engineering, \\ University of Melbourne, Parkville, VIC 3010, Australia \\ (e-mails: \{ffarokhi,ishames\}@unimelb.edu.au)}
\address[Second]{ACCESS Linnaeus Center, School of Electrical Engineering, \\ KTH Royal Institute of Technology, SE-100 44 Stockholm, Sweden (e-mail: kallej@kth.se)}

\begin{abstract} A secure and private framework for inter-agent communication and coordination is developed. This allows an agent, in our case a fleet owner, to ask questions or submit queries in an encrypted fashion using semi-homomorphic encryption. The submitted query can be about the interest of the other fleet owners for using a road at a specific time of the day, for instance, for the purpose of collaborative vehicle platooning. The other agents can then provide appropriate responses without knowing the content of the questions or the queries. Strong privacy and security guarantees are provided for the agent who is submitting the queries. It is also shown that the amount of the information that this agent can extract from the other agent is bounded. In fact, with submitting one query, a sophisticated agent can at most extract the answer to two queries. This secure communication platform is used subsequently to develop a distributed coordination mechanisms among fleet owners. 
\end{abstract} 


\end{frontmatter}

\maketitle

\section{Introduction}
The advances in communication technology has created new opportunities regarding shared economy. An example is collaborative driving, such as heavy-duty vehicle platooning or ride sharing, with the aim of reducing costs or carbon footprint of commuters or fleets~\cite{7437386}. This has motivated studies for creating appropriate incentives for the fleet owners to collaborate~\cite{farokhi2015cooperation}. Although promising, heavy-duty vehicle platooning has not yet been adopted by the larger population of vehicles. Beside technological and legal barriers, this could be partly motivated by that commercial, often competing, entities are unwilling to share their entire private data, e.g., the routes of their vehicles and their travel times, with each other even if doing so results in lower operative costs. This could be because of  privacy constraints by the customers or secretive nature of marketing agencies. Therefore, it is of interest to create private and secure match-making services for effective coordination among competing entities to facilitate these new technologies. It goes without saying that the use for such services is not limited to vehicle platooning and can be justified in many other setups, such as ride sharing, collaborative logistics, energy markets, and even online dating services. 

Following this motivation, in this paper, we create a secure and private framework for communication between two agents, fleet owners within the context of the heavy-duty vehicle platooning. In the presented framework, it is possible for an agent to ask a question or submit a simple query regarding the interest of the other agents about using a road at a specific time of the day (for forming platoons) in an encrypted fashion. This is done in such a way that the other agents can provide their responses without knowing the content of the questions or the queries. The framework is developed with the aid of semi-homomorphic encryption, which allows algebraic manipulation of the plain data without the need for decryption using appropriate computations over the encrypted data; see, e.g.,~\cite{yi2014homomorphic} about homomorphic encryption. This category of encryption techniques makes it possible for the second agent (i.e., the fleet owner receiving the encrypted question) to respond to it using appropriate manipulations of the encrypted data. Strong privacy and security guarantees are provided for the agent who submits the query. It is also shown that the amount of the information that the questioning agent can extract from all the other agents is bounded (in fact with one question a fleet owner can at most extract the answer to two questions about the interests of the other fleet owners, which could be negligible in comparison to the number of possible questions). This secure communication platform is used subsequently to develop distributed coordination mechanisms for heavy-duty vehicle platooning. Note that although the platform is developed in the context of match-making for heavy-duty vehicle platooning, the outcomes are more general and can be readily used in other examples as well (specifically if the questions or the queries are regarding selecting an element of a finite discrete set).  

This paper is in essence close to the problem of private searching on streaming data~\cite{Ostrovsky2007,yi2011private, yi2014private,Boneh2007}, where it is of interest to determine if certain important keywords are used within private encrypted messages, such as text messages and e-mails, while the content of the messages itself is not of special interest (at least not if the keywords are not utilized). However, it should be noted that the privacy guarantees of those studies are one sided (with the aim of protecting the privacy of the messages). However, in this paper, we would like to provided guarantees to both sides, i.e., both the agents posing the question and the ones responding to it.

Recently, homomorphic encryption have been utilized to solve security and privacy issues in networked control and estimation~\cite{farokhinecsys2016,kogiso2015cyber}.
Those studies are however involved with the intricacies of dynamical systems and using encryption for closing the control loop rather than creating a private/secure communication framework among multiple (possibly competing) agents. Further, they do not provide two-sided privacy guarantees (as it is not of special interest to preserve the privacy of a malicious agent if it communicates with the controller). 

The rest of the paper is organized as follows. First, the problem formulation and some necessary background material are presented in Section~\ref{sec:problem}. A secure and private communication framework between two agents is provided in Section~\ref{sec:privatecom}. Section~\ref{sec:distributed} uses the framework to construct a distributed mechanism for coordination among several fleet owners. Finally, numerical examples are presented in Section~\ref{sec:example} and the paper is concluded in Section~\ref{sec:conc}.

\section{Problem Formulation} \label{sec:problem}
In this paper, the problem formulation and the results are provided within the context of coordination among fleet owners for heavy-duty vehicle platooning. However, the results can be readily used in other classes of problems involving match-making and coordination, such as ride sharing and electricity markets. Investigating the platooning coordination problem allows us to pose concrete questions and provide meaningful privacy and security guarantees.

Consider the case with $F$ fleet owners, each owning  $I_i$, $i\in\mathcal{F}:=\{1,\dots,F\}$, heavy-duty vehicles. These vehicles can operate over various roads on a transportation network in set $\mathcal{P}$ (based on the requests of the customers of the fleet owners) and at various time intervals of the day (based on logistical constraints and customer preferences) in set $\mathcal{T}$ such that $|\mathcal{T}|<\infty$. Here, we discretize the time of the day (e.g., half an hour windows of the day) so that the number of time windows of interest is finite. We use the set of integers $\mathcal{W}:=\{1,\dots,|\mathcal{P}||\mathcal{T}|\}$ to refer to all the possible combinations of roads and time windows (in which a fleet owner might be interested or one of its heavy-duty vehicles might be operating) captured by the entries of the product set $\mathcal{P}\times\mathcal{T}$. In fact, it can be said that the set $\mathcal{W}$ is isomorph to the set $\mathcal{P}\times\mathcal{T}$. Our goal is to develop a secure and private communication framework for the fleet owners to identify potential heavy-duty vehicles that can form platoons. This is clearly possible if the fleet owners exchange the time and the roads over which all their heavy-duty vehicles operate; however, that would violate the privacy of the customers and the fleet owners (possibly jeopardizing their competitive advantages). Therefore, we would like to create a communication platform so that the fleet owners can only enquire about each other's interests and logistical constraints under strict privacy constraints. In fact, we show that the enquiring fleet owner does not leak any private information (i.e., the other fleet owners cannot realize the road and the time window of interest of the enquiring fleet owner). In addition, we show that even with most sophisticated manoeuvres the enquiring fleet owner can only extract information on the interests of the other fleet owners about at most two pairs of roads and time windows (which considering the sheer number of roads and time windows is negligible). To do so, we use the Paillier's encryption method, which is a semi-homomorphic encryption technique. The encryption method is introduced in the following subsection.

\subsection{Background Material}
In this subsection, the Paillier's encryption technique is briefly introduced~\cite{Paillier1999}.
The method (or rather its security) relies on the Decisional Composite Residuosity Assumption, i.e., for given integers $N\in\mathbb{Z}$ and $x\in\mathbb{Z}_{N^2}$, it is ``hard'' to decide whether there exists $y\in\mathbb{Z}_{N^2}$ such that $x\equiv y^N \modd N$. Here, the notation $\mathbb{Z}_{N}$ denotes the set of integers modulo $N$ for all~$N\in\mathbb{N}$. More information regarding the assumption can be found in~\cite{Paillier1999,yi2014homomorphic}.
The encryption scheme is as follows. First the public and private keys are generated. To do so, large prime numbers $p$ and $q$ are selected randomly and independently of each other such that $\gcd(pq,(1-p)(1-q))=1$, where $\gcd(a,b)$ refers to the greatest common divisor of $a$ and $b$. The public key (which is shared with all the parties and is used for encryption) is $N=pq$. The private key (which is only available to the entity that needs to decrypt the data) is $(\lambda,\mu)$ with $\lambda=\lcm(p-1,q-1)$ and $\mu=\lambda^{-1}\modd N$, where $\lcm(a,b)$ is the least common multiple of $a$ and $b$. The ciphertext of a plain message $t\in\mathbb{Z}_N$ can be constructed using $E(t;r)=(N+1)^tr^N\modd N^2$, where $r$ is randomly selected with uniform probability from $\mathbb{Z}^*_N:=\{x\in\mathbb{Z}_N|\gcd(x,N)=1\}$. Finally, to decrypt any ciphertext $c\in\mathbb{Z}_{N^2}$, we may use $D(c)=L(c^\lambda \modd N^2)\mu\modd N$, where $L(x)=(x-1)/N$. The correctness of the Paillier's encryption technique implies that $D(E(t;r))=t$ for all $r\in\mathbb{Z}^*_N$ and all $t\in\mathbb{Z}_N$~\cite{Paillier1999}.
The following important property shows that the Paillier's encryption is a semi-homomorphic encryption scheme, i.e., algebraic manipulation of the plain data is possible without the need for decryption using appropriate computations over the encrypted data.

\begin{proposition} \label{prop:cipher} The following identities hold:
\begin{enumerate}
\item For all $r,r'\in\mathbb{Z}^*_N $ and $t,t' \in\mathbb{Z}_N$ such that $t+t'\in\mathbb{Z}_N$, $E(t;r)E(t';r')\modd N^2=E(t+t';rr')$;
\item For all $r\in\mathbb{Z}^*_N$ and $t,t' \in\mathbb{Z}_N$ such that $tt'\in\mathbb{Z}_N$, $E(t;r)^{t'}\modd N^2=E(t't;r^{t'})$.
\end{enumerate}
\end{proposition}

\begin{pf} The proof follows from simple algebraic manipulations and can be found in~\cite{Paillier1999}. \hfill $\square$
\end{pf}

Proposition~\ref{prop:cipher} shows that summation and multiplication can be performed on the encrypted data. This is used subsequently for creating a secure and private method for coordination between fleet owners for heavy-duty vehicle platooning.

\section{Secure and Private Communication Framework} \label{sec:privatecom}
In this section, we restrict ourselves to two fleet owners. We subsequently generalize the setup to develop a distributed coordination mechanism. Let us assume that the first fleet owner would like to know if the second fleet owner has any heavy-duty vehicle that uses the path and the time of the day associated with $w\in\mathcal{W}$. However, it does not wish to let the second fleet owner to know $w$ perhaps due to the competitive nature of the fleet owners or the privacy constraints imposed by the end user. Therefore, the first fleet owner construct an encrypted vector $x\in\mathbb{Z}^{|\mathcal{W}|}$ such that the $i$-th element of $x$ is given by
\begin{align*}
x_i=
\begin{cases}
E(1;r_i), & i=w,\\
E(0;r_i), & \mbox{otherwise}.
\end{cases}
\end{align*}
Note that the presence of the random element $r_i$ ensures that with a high probability cyphertexts associated with $0$ are different\footnote{In fact, for industry standards that assume both $q$ and $q$ are of the order of 1024 bits, we have $N=\mathcal{O}(2^{2048})$. Thus the probability of selecting the same $r$ twice even in a vector of millions of elements is smaller than $10^{-1000}$, which is practically zero.}. Then the first fleet owner transmits $x$ to the second fleet owner. The second fleet owner computes the set $\overline{\mathcal{W}}\subseteq\mathcal{W}$ as the set of all $w'\in \mathcal{W}$ with which at least one of its heavy-duty vehicles is associated. The second fleet owner then computes
\begin{align*}
y=\bigg(\prod_{j\in\overline{\mathcal{W}}} x_j^{v_j}\modd N^2\bigg)\modd N^2,
\end{align*}
where $v_j$ is randomly selected from the set $\{1,\dots,N-1\}$ with a uniform probability. 
We summarize the procedures of the first and the second fleet owner in Algorithms~\ref{alg:1} and~\ref{alg:2}.
We can prove the following useful result pointing to the correctness of our proposed methodology.

\begin{algorithm}[t]
\caption{\label{alg:1} Procedure {\sc SubmitQuery} for the first fleet owner.}
\begin{algorithmic}[1]
\Procedure{SubmitQuery}{$w$} 
\State \# Computed by the first fleet owner
\For{$i\in\mathcal{W}$}
\If{$i=w$}
\State $x_i\leftarrow E(1;r_i)$
\Else
\State $x_i\leftarrow E(0;r_i)$
\EndIf
\EndFor
\State \textbf{return} $x$ 
\EndProcedure
\end{algorithmic}
\end{algorithm}

\begin{algorithm}[t]
\caption{\label{alg:2} Procedure {\sc ReturnResponse} for the first fleet owner.}
\begin{algorithmic}[1]
\Procedure{SubmitQuery}{$x$,$\overline{\mathcal{W}}$} 
\State \# Computed by the second fleet owner
\State $y\leftarrow 1$
\For{$i\in\overline{\mathcal{W}}$}
\State Select $v_i$ randomly from $\{1,\dots,N-1\}$
\State $y\leftarrow y(x_i^{v_i}\modd N^2)\modd N^2$
\EndFor
\State \textbf{return} $y$ 
\EndProcedure
\end{algorithmic}
\end{algorithm}

\begin{proposition} If the first and the second fleet owner, respectively, use Algorithms~\ref{alg:1} and~\ref{alg:2}, then $D(y)\neq 0$ if any of the heavy-duty vehicles owned by the second fleet owner uses the path and time window associated with $w$ and $D(y)=0$ otherwise. 
\end{proposition}

\begin{pf} The proof follows from the construction of the vector $x$ and the application of Proposition~\ref{prop:cipher}.\hfill$\square$
\end{pf}

To be able to assess the security and privacy of the proposed method from the perspective of the first fleet owner, we present the following definition. This notion is derived from what is known in the encryption literature as semantic security. To present this definition, we need to define the notion of negligible functions. A function $f:\mathbb{N}\rightarrow\mathbb{R}_{\geq 0}$ is called negligible if, for any $c\in\mathbb{N}$, there exists $n_c\in\mathbb{N}$ such that $f(n)\leq 1/n^c$ for all $n\geq n_c$~\cite{Ostrovsky2007}.

\begin{definition} \label{def:secure} Let the second fleet owner propose $w_1,w_2\in\mathcal{W}$. The first fleet owner chooses at random $w$ from $\{w_1,w_2\}$ with equal probability and sends $x$ constructed using Algorithm~\ref{alg:1}. The second fleet owner can based on its knowledge of $x$ guess $w$. This guess is denoted by $w'$. The second fleet owner's advantage\footnote{The advantage refers to its superior performance to that of a ``dart throwing monkey.''} is given by $\mathrm{Adv}(k):=|\mathbb{P}\{w=w'\}-1/2|$, where $k$ denotes the security parameter, e.g., $\min(p,q)$ in the Paillier's technique. The proposed strategy is defined to be secure and private if $\mathrm{Adv}$ is negligible. 
\end{definition}

Definition~\ref{def:secure}, although long and cumbersome, says that the proposed strategy is  secure and private if, essentially, the performance of the second fleet owner in guessing the first fleet owner's preference $w$ is not better than a pure random number generator. 

\begin{proposition} \label{prop:privacyfirst} 
Under the Decisional Composite Residuosity Assumption, Algorithm~\ref{alg:1} is secure and private in the sense of Definition~\ref{def:secure}.
\end{proposition}

\begin{pf} The proof follows from the application of the results of~\cite{Paillier1999}.\hfill$\square$
\end{pf}

Unfortunately, the privacy and security guarantees are weaker for the second fleet owner. This is because by definition if the first and the second fleet owner, respectively, use Algorithms~\ref{alg:1} and~\ref{alg:2}, the first fleet owner can successfully determine if the second fleet owner is in possession of any heavy-duty vehicles that operate on the path and in the time window associated with $w$ (therefore some private information is leaked even in the best of situations). However, one might be able to argue that the first fleet owner can potentially extract more information by not following Algorithm~\ref{alg:1}. This is in fact true (pointing to a deeper erosion of privacy). However, in what follows, we prove that the amount of  information the first fleet owner can optimally extract is limited (with the bound being extremely small). 
Assume that the first fleet owner does not use Algorithm~\ref{alg:1}. Instead, it constructs the vector $x$ such that $x_i=E(\tilde{x}_i;r_i)$ for some integer $\tilde{x}_i\in\mathbb{Z}_N$. We can prove the following lemma.

\begin{lemma} \label{lem:1} If, for all $i$, $x_i=E(\tilde{x}_i;r_i)$ for some integer $\tilde{x}_i\in\mathbb{Z}_N$, then
\begin{align}
D(y)=\bigg(\sum_{w\in\mathcal{W}} \tilde{x}_iv_iz_i\bigg)\modd N,
\end{align}
where $z_i=1$ if any of the heavy-duty vehicles owned by the second fleet owner uses the path and the time window associated with $i\in\mathcal{W}$ and $z_i$ otherwise.
\end{lemma}

\begin{pf} The proof follows from the construction of the vector $x$, $\tilde{x}_i,\forall i$, and the application of Proposition~\ref{prop:cipher}.\hfill$\square$
\end{pf}

Following Lemma~\ref{lem:1}, the first fleet owner needs to solve $D(y)=\sum_{w\in\mathcal{W}} \tilde{x}_iv_iz_i \modd N$ to find $z_i$ for all $i$. Note that the first fleet owner can introduce the change of variable $\xi_i=v_iz_i$ and instead solve the linear equation modulo $D(y)=\sum_{w\in\mathcal{W}} \tilde{x}_i\xi_i \modd N$. Evidently, $z_i=1$ if $\xi_i\neq 0$ and $z_i=0$ otherwise. Let us form the set
\begin{align}
\Xi:=\left\{\xi\in\mathbb{Z}_N^{|\mathcal{W}|}\,\big|\,D(y)=\bigg(\sum_{w\in\mathcal{W}} \tilde{x}_i\xi_i\bigg) \modd N\right\},
\end{align}
which represents the set of all solutions of the linear equation modulo $D(y)=\sum_{w\in\mathcal{W}} \tilde{x}_i\xi_i \modd N$. We can prove the following useful result.

\begin{proposition} \label{prop:numberofsol} Let $t=|\{i\,|\,\tilde{x}_i\neq 0\}|>1$. Then $|\Xi|\geq (N-1)^{t-1}$ if there exists $i$ such that $\gcd(\tilde{x}_i,N)=1$. 
\end{proposition}

\begin{pf} For $i$ such that $\gcd(\tilde{x}_i,N)=1$, there exists $\tilde{x}_i^{-1}\modd N$. Thus we get
\begin{align*}
\xi_i=\bigg(D(y)-\sum_{j\neq i}\tilde{x}_i^{-1}\tilde{x}_j\xi_j\bigg)\modd N.
\end{align*}
Therefore, all $(\xi_j)_{j\neq i}$ are free variables, i.e., for any selection of $(\xi_j)_{j\neq i}$, there exists $\xi_i$ that satisfies the linear equation modulo $D(y)=\sum_{w\in\mathcal{W}} \tilde{x}_i\xi_i \modd N$. This points to that the number of solutions of the linear equation modulo (which is equal to $|\Xi|$) is equal to the number of all the possible choices of $(\xi_j)_{j\neq i}$.\hfill$\square$
\end{pf}

This shows that even if only two $\tilde{x}_i$ are non-zero, the number of possible solutions of $D(y)=\sum_{w\in\mathcal{W}} \tilde{x}_i\xi_i \modd N$, i.e., $|\Xi|$, is larger than $N-1$. The situation worsens as more $\tilde{x}_i$ become non-zero because $t$ (the number of the non-zero elements) appear as an exponent. Noting that the encryption relies on $N$ being extremely large\footnote{Routinely, $p,q$ are selected as prime numbers with the length of 1024 bits pointing to that $N=\mathcal{O}(2^{2048})$.}, the first fleet owner needs to check a huge number of possible solutions
, which is numerically impractical. 

\begin{proposition} \label{prop:numberofsol2} Let $t=|\{i\,|\,\tilde{x}_i\neq 0\}|>2$. Then $|\Xi|\geq 2(N-1)^{t-2}$ if there does not exist $i$ such that $\gcd(\tilde{x}_i,N)=1$. 
\end{proposition}

\begin{pf} If there does not exist $i$ such that $\gcd(\tilde{x}_i,N)=1$, we can construct two sets where in the first one $\tilde{x}_i$ is divisible by $q$ and in the second one $\tilde{x}_i$ is divisible by $p$ (note that $\tilde{x}_i$ cannot be divisible by both as otherwise it will be larger than $\lcm(p,q)=pq=N$). Let the sets be denoted by $\mathcal{J}_1$ and $\mathcal{J}_2$, respectively. In this case, we can write
\begin{align*}
D(y)=q\bigg(\sum_{j\in\mathcal{J}_1} \xi_j\underbrace{\bigg(\frac{\tilde{x}_j}{q}\bigg)}_{\tilde{x}'_j}\bigg)
+p\bigg(\sum_{j\in\mathcal{J}_2} \xi_j\underbrace{\bigg(\frac{\tilde{x}_j}{p}\bigg)}_{\tilde{x}'_j}\bigg) \modd N.
\end{align*}
Noting that $\gcd(p,q)=1$ (since $p$ and $q$ are prime numbers), this equation can be separated into 
\begin{subequations}
\begin{align}
\alpha&=\sum_{j\in\mathcal{J}_1} \xi_j\tilde{x}'_j \modd N,\label{eqn:proof:1}\\
\beta&=\sum_{j\in\mathcal{J}_2} \xi_j\tilde{x}'_j \modd N,\label{eqn:proof:2}
\end{align}
\end{subequations}
where $\alpha=D(y)\bar{\alpha}$ and $\beta=D(y)\bar{\beta}$ with $\bar{\alpha}$ and $\bar{\beta}$ denoting B\'{e}zout coefficients, i.e., $\bar{\alpha} q+\bar{\beta}p=1$. There are only two B\'{e}zout coefficients that satisfy $|\bar{\alpha}|<p$ and $|\bar{\beta}|<q$~\cite[Proposition\,13, p.\,60]{vialar2015handbook}. The number of the solutions of~\eqref{eqn:proof:1} can be lower bounded with the same line of reasoning as in Proposition~\ref{prop:numberofsol} by $(N-1)^{|\mathcal{J}_1|-1}$. Similarly, the number of the solutions of~\eqref{eqn:proof:2} can be lower bounded by $(N-1)^{|\mathcal{J}_2|-1}$. This concludes the proof.\hfill$\square$
\end{pf}

Proposition~\ref{prop:numberofsol2} shows that, by smart planning and sophisticated manoeuvres, the first fleet owner can realize if the second fleet owner has any heavy-duty vehicles on the roads and the time windows associated with two entries of $\mathcal{W}$ instead of one by following Algorithm~\ref{alg:1}. Assuming that $|\mathcal{W}|$ is large, this might not matter in practice. 

\begin{remark}[Computational Complexity] Before moving \linebreak to more complex communication 
structures, the computational complexity of the proposed algorithms should be discussed. Algorithm~\ref{alg:1} involves $|\mathcal{W}|$ encryption operations. Each encryption operation has a cost that is a (nonlinear) function of $N$. Noting that the size of $N$ is often a constant set by standards of the industry, the cost of the encryption is also constant (albeit a large one). Therefore, the  computational complexity of Algorithm~\ref{alg:1} is $\mathcal{O}(|\mathcal{W}|)$. Algorithm~\ref{alg:2}, on the other hand, involves $|\overline{\mathcal{W}}|\leq \max_{i} I_i$ exponentiations and multiplications. Therefore, the computational complexity of Algorithm~\ref{alg:2} is $\mathcal{O}(\max_{i} I_i)$. Therefore, the total computational complexity of Algorithms~\ref{alg:1} and~\ref{alg:2} is $\mathcal{O}(|\mathcal{W}|)$ because, in practice, $\max_{i} I_i\ll |\mathcal{W}|$. Finally, since the first fleet owner needs to submit $\mathcal{O}(\max_{i} I_i)$ questions or queries to solve the platooning coordination problem, the computational complexity of the whole task is $\mathcal{O}(|\mathcal{W}|\max_{i} I_i)$.
\end{remark}

\section{Distributed Coordination}
\label{sec:distributed}
Now, we use the results of the previous section to develop a distributed mechanism for the fleet owners to coordinate their efforts. Let the undirected graph $\mathcal{G}$ with the vertex set $\mathcal{F}$ (i.e., the vertices are the fleet owners) and the edge set $\mathcal{E}\subseteq\mathcal{F}\times\mathcal{F}$ capture the communication structure among the agents. A walk over $\mathcal{G}$ (not to be mistaken with roads over which the heavy-duty vehicles travel on real transportation network) is a sequence of (not necessarily unique) vertices $\mathcal{L}=(v_0,\dots,v_k)$ such that $(v_i,v_{i+1})\in\mathcal{E}$ for all $0\leq i\leq k-1$. 
A $k$-connected graph is  a graph that after removing any $k$ vertices (and all the edges connected to those vertices) is still a connected graph. We make the following standing assumption.

\begin{assumption} \label{assum:1} $\mathcal{G}$ is $2$-connected.
\end{assumption}

Assumption~\ref{assum:1} states that even if one of the fleet owners is removed,  all the remaining ones can still communicate with each other successfully. This allows us to develop an algorithm for the fleet owners to collaboratively respond to queries by avoiding communication with fleet owners that have submitted the queries. Let us consider the case where fleet owner $\ell\in\mathcal{F}$ would like to find out if there exists any other fleet owner such that one of its heavy-duty vehicles operate over the path and the time window associated with $w\in\mathcal{W}$. Further, let there be a walk $\mathcal{L}=(v_0,\dots,v_k)$ over $\mathcal{G}$ such that $v_0=v_k=\ell$ (therefore the walk is a loop) while $v_j\neq \ell$ for all $1\leq j\leq k-1$. Existence of such a walk is guaranteed by Assumption~\ref{assum:1}. Similar, to the previous section, fleet owner $i$ can follow Algorithm~\ref{alg:1} to construct the vector $x$, thus submitting an encrypted query for coordination. After that all the other fleet owners in the walk $\mathcal{L}$ can follow the procedure in Algorithm~\ref{alg:3} to respond to the query of fleet owner $\ell$. In this algorithm, $\overline{\mathcal{W}_j}\subseteq\mathcal{W}$ denotes the set of all $w\in \mathcal{W}$ with which at least one of its heavy-duty vehicles of fleet owner $j\in\mathcal{F}$ is associated. The following proposition shows that if fleet owner $\ell$ uses Algorithm~\ref{alg:1} and all the fleet owners in $\mathcal{L}\setminus\{\ell\}$ use Algorithm~\ref{alg:3}, the provided response is correct.

\begin{algorithm}[t]
\caption{\label{alg:3} Procedure {\sc DistResponse} for the fleet owners in the walk $\mathcal{L}$ responding to the query of the fleet owner $\ell$ distributedly.}
\begin{algorithmic}[1]
\Procedure{DistResponse}{$x$,$\mathcal{L}$,$(\overline{\mathcal{W}_j})_{j\in\mathcal{L}\setminus\{\ell\}}$} 
\State \# Computed by the fleet owners in $\mathcal{L}$ except $i$
\For{$j=v_1,\dots,v_{k-1}$}
\For{$i\in\overline{\mathcal{W}}_j$}
\State Select $\omega_i$ randomly in $\{1,\dots,\lfloor N/(|\mathcal{L}|-2)\rfloor \}$
\State $x_i\leftarrow x_i^{\omega_i}\modd N^2$
\EndFor
\EndFor
\State $y\leftarrow 1$
\For{$i\in\overline{\mathcal{W}}_{v_k}$}
\State Select $\omega_i$ randomly in $\{1,\dots,\lfloor N/(|\mathcal{L}|-2)\rfloor \}$
\State $y\leftarrow y(x_i^{\omega_i}\modd N^2)\modd N^2$
\EndFor
\State \textbf{return} $y$ 
\EndProcedure
\end{algorithmic}
\end{algorithm}

\begin{proposition} If fleet owner $\ell$ uses Algorithm~\ref{alg:1} and all the fleet owners in $\mathcal{L}\setminus\{\ell\}$ use Algorithm~\ref{alg:3}, then $D(y)\neq 0$ if any of the heavy-duty vehicles owned by the fleet owner in $\mathcal{L}\setminus\{\ell\}$ uses the path and time window associated with $w$, and $D(y)=0$ otherwise. 
\end{proposition}

\begin{pf} The proof follows from the application of Proposition~\ref{prop:cipher}.\hfill$\square$
\end{pf}

Note that, if fleet owner $\ell$ is interested in figuring out the possibility of forming a platoon with all the other fleet owners (and not a select few), it should find a walk $\mathcal{L}$ that spans all the vertices of the graph.

A similar result as in Proposition~\ref{prop:privacyfirst} can be proved for the enquiring fleet owner in this case as well. Therefore, we focus on the privacy guarantees of the other fleet owners in $\mathcal{L}\setminus\{\ell\}$. Now, we can prove the following lemma.

\begin{lemma} \label{lem:2} If, for all $i$, $x_i=E(\tilde{x}_i;r_i)$ for some integer $\tilde{x}_i\in\mathbb{Z}_N$, then
\begin{align}
D(y)=\bigg(\sum_{w\in\mathcal{W}} \tilde{x}_iv_i\bigg(\sum_{j\in\mathcal{L}\setminus\{\ell\}}z_i^j\bigg)\bigg)\modd N,
\end{align}
where $z_i^j=1$ if any of the heavy-duty vehicles owned by fleet owner $j\in\mathcal{L}\setminus\{\ell\}$ uses the path and the time window associated with $i\in\mathcal{W}$ and $z_i^j=0$ otherwise.
\end{lemma}

\begin{pf} The proof is similar to that of Lemma~\ref{lem:1}.\hfill$\square$
\end{pf}

Similarly, following Lemma~\ref{lem:2}, the enquiring fleet owner $\ell$ must solve the linear equation modulo 
\begin{align*}
D(y)=\bigg(\sum_{w\in\mathcal{W}}\sum_{j\in\mathcal{L}\setminus\{\ell\}} \tilde{x}_i\xi_i^j \bigg)\modd N,
\end{align*}
where $z_i^j=1$ if $\xi_i^j\neq 0$ and $z_i^j=0$ otherwise. Let us construct the set of all possibilities
\begin{align}
\Xi:=\bigg\{&(\xi_i^j)_{j\in\mathcal{L}\setminus\{\ell\}}\in\mathbb{Z}_N^{|\mathcal{W}|(|\mathcal{L}|-2)}\,\big|\,\nonumber\\
&\hspace{.3in}D(y)=\bigg(\sum_{w\in\mathcal{W}}\sum_{j\in\mathcal{L}\setminus\{\ell\}} \tilde{x}_i\xi_i^j \bigg)\modd N\bigg\}. 
\end{align}
We can prove the following useful result regarding the size of the set $\Xi$.

\begin{proposition} \label{prop:numberofsoldistributed} 
The following two statements hold:
\begin{itemize}
\item Let $t=|\{i\,|\,\tilde{x}_i\neq 0\}|>1$. Then $|\Xi|\geq (|\mathcal{L}|-2)(N-1)^{t-1}$ if there exists $i$ such that $\gcd(\tilde{x}_i,N)=1$.
\item Let $t=|\{i\,|\,\tilde{x}_i\neq 0\}|>2$. Then $|\Xi|\geq 2(|\mathcal{L}|-2)^2(N-1)^{t-2}$ if there does not exist $i$ such that $\gcd(\tilde{x}_i,N)=1$.
\end{itemize} 
\end{proposition}

\begin{pf} The proof follows a similar line of reasoning as in Propositions~\ref{prop:numberofsol} and~\ref{prop:numberofsol2}.
\end{pf}

This proposition shows that the privacy guarantees of the fleet owners of the walk is stronger than those in the case of two agents as the responses of all the agents gets mixed. Therefore, even if the fleet owner can extract the aggregate answers to two questions, it would not know which one of the fleet owners from the set $\mathcal{L}\setminus\{\ell\}$ has responded positively. In practice, however, when the fleet owners show up for forming a platoon the set of all possibilities gets further narrowed down.


\section{Numerical Example}
\label{sec:example}

\begin{figure}
\includegraphics[width=1\linewidth]{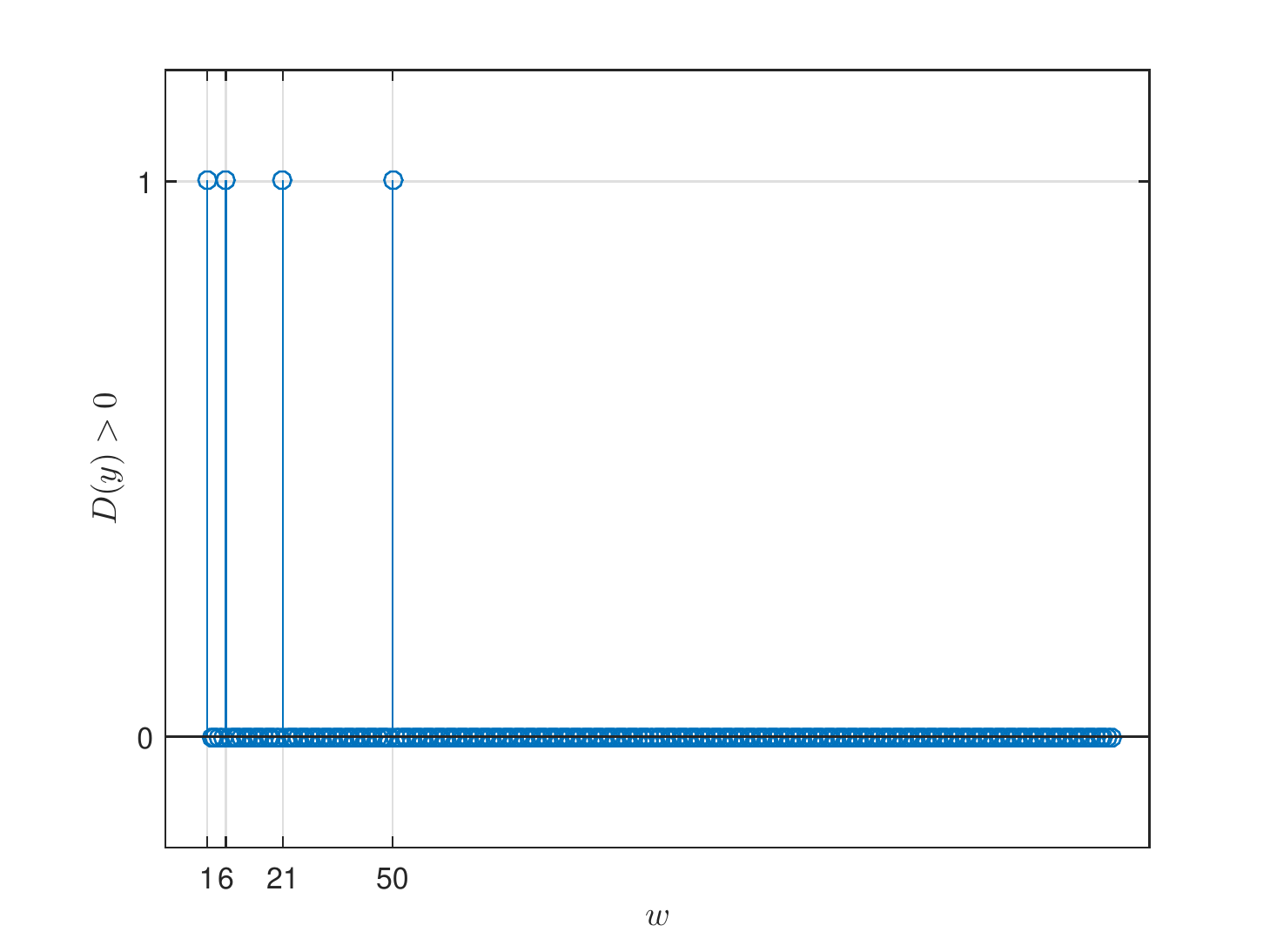}
\caption{\label{fig:2} The outcome of Algorithms~\ref{alg:1} and~\ref{alg:2} for various $w$.}
\end{figure}

\begin{figure}
\includegraphics[width=1\linewidth]{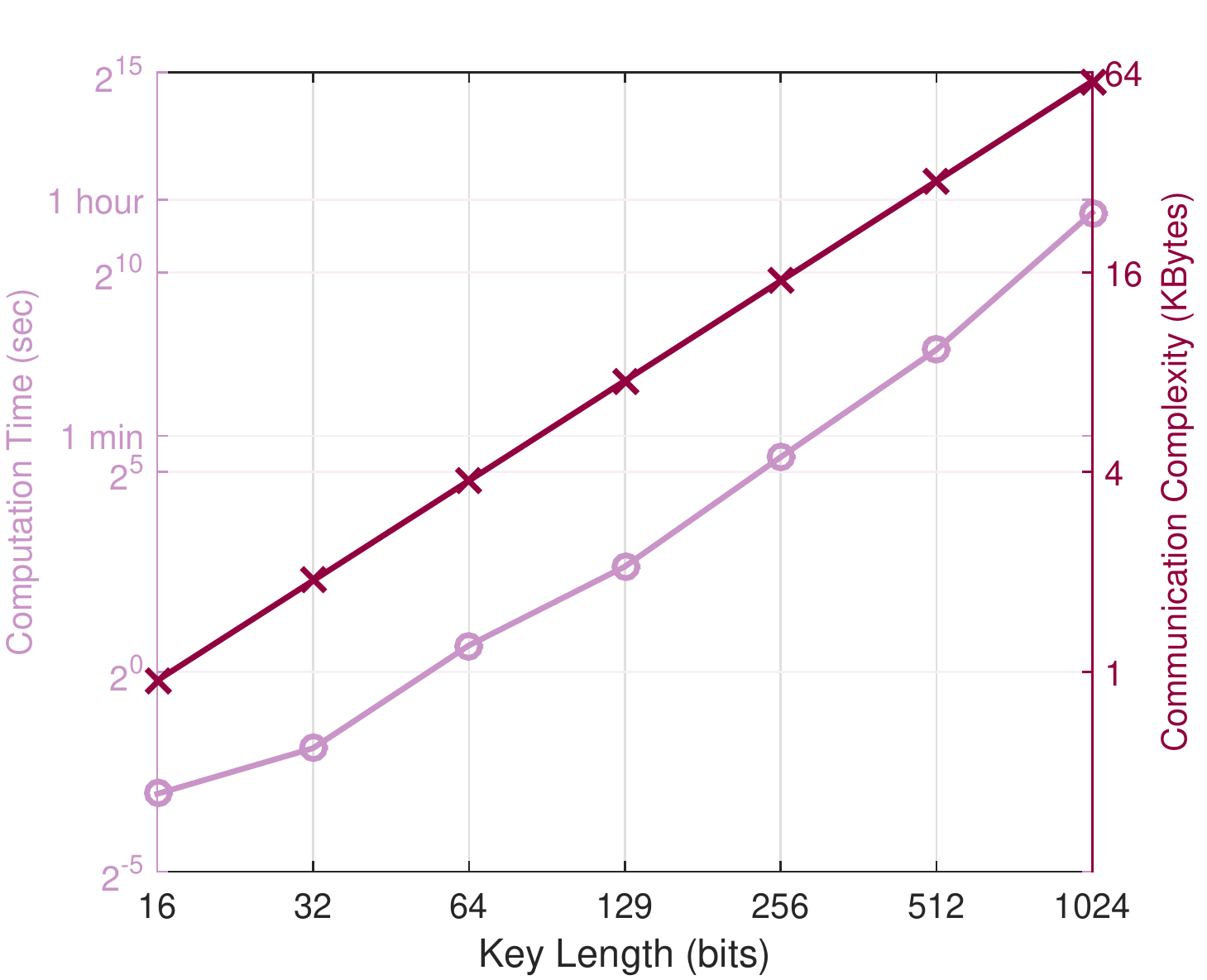}
\caption{\label{fig:1} The computation time (\comp) and the communication burden (\comm) associated with executing Algorithms~\ref{alg:1} and~\ref{alg:2} versus the key length.}
\end{figure}

In this section, we review some practical aspects of the developed framework. Specifically, we review the  computation time and the communication burden associated with executing the proposed algorithms. 

Let us consider an example, where two fleet owners need to coordinate their efforts for organizing heavy-duty vehicle platoons. Assume that there are $|\mathcal{P}|=10$ roads and time of the day is discretized into $|\mathcal{T}|=24$ one-hour windows. Therefore, $|\mathcal{W}|=240$. The second fleet owner has heavy-duty vehicles on roads and time windows associated with $w=1,6,21,50$. Figure~\ref{fig:2} illustrates the outcome of Algorithms~\ref{alg:1} and~\ref{alg:2} for various queries submitted by the first fleet owner with a key length of 128 bits. The vertical axis of Figure~\ref{fig:2} is equal to one if $D(y)>0$ and is equal to zero otherwise. Evidently, $D(y)>0$ for only $w=1,6,21,50$. Therefore, as expected, upon following the proposed algorithms, the second fleet owner can correctly respond to the query of the first fleet owner without even knowing the content of the query. However, this secure communication channel comes at a price. Figure~\ref{fig:1} shows the computation time and the communication burden associated with executing Algorithms~\ref{alg:1} and~\ref{alg:2} versus the key length, measured in bits. The computation is done with Python programming language on Windows 7 over a PC with Intel(R) i7-4770 CPU at 3.40GHz and 16GB of RAM. The computation time rapidly increases with increasing the key length. The amount of  data the fleet owners need to communicate also increases with the key length. Note that the security of the encryption is related to the key length. In fact, the computational complexity of a brute-force attack that requires the test of all the keys of a specific length grows exponentially with the key length. Note that the computational complexity grows polynomially with the key length (the linear appearance is due to the logarithmic scaling of both axes in Figure~\ref{fig:1}). In fact, upon fitting an appropriate curve, we get that the computational time (in sec) grows as  $\mathcal{O}(k^{2.44})$ with $k$ denoting the key length. On the other hand, the computational burden is a linear function of the key length (the slope of the line in Figure~\ref{fig:1} is equal to one) as the size of the integers that needs to be transmitted grows linearly with the key length.

\section{Conclusions and Future Work}
\label{sec:conc}
A secure and private framework for communication between two fleet owners was proposed. This secure communication platform was generalized to create distributed coordination mechanisms among fleet owners for  heavy-duty vehicle platooning. The future work will focus on developing a centralized coordination mechanism for reducing the computational complexity or outsourcing burden to cloud computing services as well as the application of the framework to other services.

\bibliographystyle{IEEEtran}
\bibliography{citation}

\end{document}